\documentclass{article}
\usepackage{ijcai23}

\usepackage{times}
\usepackage{soul}
\usepackage{url}
\usepackage[colorlinks=true,linkcolor=blue,citecolor=blue]{hyperref}
\usepackage[utf8]{inputenc}
\usepackage[small]{caption}
\usepackage{graphicx}
\usepackage{amsmath}
\usepackage{amsthm}
\usepackage{booktabs}
\usepackage{algorithm}
\usepackage{algorithmic}
\usepackage[switch]{lineno}
\usepackage{beramono}
\usepackage[T1]{fontenc}
\usepackage{xcolor}

\newcommand{\citet}[1]{\cite{#1}}

\definecolor{mdgreen}{rgb}{0.05,0.6,0.05}
\definecolor{dred}{rgb}{0.6,0.05,0.05}
\newif\ifshowcomments
\showcommentstrue

\usepackage{cleveref}
\crefformat{section}{\S#2#1#3}
\crefformat{subsection}{\S#2#1#3}
\crefformat{subsubsection}{\S#2#1#3}

\usepackage{times}
\usepackage{latexsym}
\usepackage{multicol}
\usepackage{multirow}
\usepackage{hyperref}
\usepackage{booktabs}
\usepackage{graphicx}
\usepackage{enumitem}
\usepackage{xcolor}
\definecolor{lightblue}{RGB}{0,127,255}
\usepackage{amsmath}

\newcommand{\prob}{\text{I\kern-0.1em P}}

\usepackage{comment}

\title{Users are the North Star for AI Transparency}

\author{
Alex Mei\thanks{Equal contribution}$^1$\and
Michael Saxon$^{*1}$
\and
Shiyu Chang$^{1}$\and
Zachary C. Lipton$^2$\and
William Yang Wang$^1$
\affiliations
$^1$Department of Computer Science, University of California, Santa Barbara\\
$^2$Department of Computer Science, Carnegie Mellon University\\
\emails
\texttt{\small \{alexmei, chang87, william\}@cs.ucsb.edu, saxon@ucsb.edu, zlipton@cmu.edu} 
}

\begin{document}

\maketitle

\begin{abstract}
Despite widespread calls for \emph{transparent} artificial intelligence systems, the term is too overburdened with disparate meanings to express precise policy aims or to orient concrete lines of research. Consequently, stakeholders often talk past each other, with policymakers expressing vague demands and practitioners devising solutions that may not address the underlying concerns. Part of why this happens is that a clear ideal of AI transparency goes unsaid in this body of work. We explicitly name such a \emph{north star}---transparency that is user-centered, user-appropriate, and honest. We conduct a broad literature survey, identifying many clusters of similar conceptions of transparency, tying each back to our north star with analysis of how it furthers or hinders our ideal AI transparency goals. We conclude with a discussion on common threads across all the clusters, to provide clearer common language whereby policymakers, stakeholders, and practitioners can communicate concrete demands and deliver appropriate solutions. We hope for future work on AI transparency that further advances confident, user-beneficial goals
and provides clarity to regulators and developers alike.
\end{abstract}

\section{Introduction}\label{sec:intro}

The discourse surrounding the societal impacts
of artificial intelligence (AI) systems abounds with 
calls, both in popular demands and formal regulations, 
for greater \emph{transparency}.
Sometimes these demands invoke the word transparency directly,
while other cases invoke similarly vague surrogates
like ``meaningful information" \cite{gdpr}.
However, the term is too overloaded with distinct meanings
to express concrete policy objectives or technical claims alone \cite{andrada2022varieties}.
The term is a prototypical example of AI's suitcase words
\cite{lipton2018troubling}.
Although this breadth can be valuable in uniting
members of disparate research communities 
toward high-level desiderata,
concrete aims and advances 
must be expressed in more precise language. 
Unfortunately, researchers, corporations, journalists, 
regulators, and members of the general public 
often invoke \emph{transparency}
in contexts where greater precision is required
and consequently, talk past each other.

\begin{table*}[t!]
\centering
\resizebox{\linewidth}{!}{
\begin{tabular}{ l p{1.2\linewidth}} 
 \toprule
 \textbf{Perspective} & \textbf{Definition of \textit{transparency}} \\ 
  
  \midrule
  Public Policy & Any meaningful information relating to consumer data is disclosed in comprehensible language \cite{gdpr,ec2019ethics}. \\
  
  \midrule
  Data Collection & Disclosure of collection methods and privacy policies in a consumer-understandable manner \cite{driscoll2014big,agozie2021discerning}. \\
  
  \midrule
  Data Processing & Comprehensible disclosure of methods in which consumer data is processed, stored, and used \cite{kirrane2021specialk}.  \\ 

  \midrule
  Reproducibility & Disclosure of important information to reproduce a system's performance \cite{gundersen2018state} \\ 
  
  \midrule
  Intelligibility & Disclosure of pertinent system functionality and limitations comprehensible to stakeholders \cite{vaughan2020human,ehsan2021expanding}. \\

  \midrule
  Interpretability & Explanation that aids understanding of system functionality \cite{lipton2018mythos,watson2019addressing}. \\
  
  \midrule
  Fairness & Disclosure regarding representation and treatment to ensure equity among groups \cite{castillo2019fairness,bhatt2021uncertainty}. \\ 
   
  \bottomrule
\end{tabular}
}
\vspace{-1ex}
\caption{Seven examples of how \emph{transparency} can be defined from different perspectives, with citations containing usage as such.}
\label{table:definitions}
\end{table*}

Depending on the context, researchers may invoke transparency
in connection with data collection
\cite{driscoll2014big,agozie2021discerning},
data processing \cite{kirrane2021specialk},
interpretable systems \cite{lipton2018mythos,watson2019addressing},
or fairness issues \cite{castillo2019fairness},
among other concerns (Table \ref{table:definitions}).
Even in European Union (EU) regulations,
which pioneered global AI policy, particularly
the General Data Protection Regulation (GDPR) \cite{gdpr},
and the Ethics Guidelines for Trustworthy AI \cite{ec2019ethics},
the vague demands for ``meaningful information''
and ``comprehensible language'' 
have forced legal scholars
and AI practitioners to speculate the precise meaning of \emph{transparency}
\cite{wachter2017counterfactual,selbst2018meaningful}.
Can these disparate research threads be unified to advance a 
coherent vision for improved AI transparency?

We believe that \textbf{ideal AI transparency gives users and stakeholders the tools to rationally, autonomously, and confidently decide for themselves whether an AI system and its decisions are trustworthy}. In particular, this means explanations or descriptions that are \emph{user-appropriate}, \emph{user-centered}, 
and \emph{honest}. 
We define these attributes as follows. 

\begin{itemize}[leftmargin=*]
\setlength\itemsep{0em}
\setlength\topsep{0em}
    \item \textbf{User-appropriate}: information conveyed to a stakeholder is understandable in content, style, and level of detail
    \item \textbf{User-centered}: insightful regarding the behaviors observed by a user in their own interactions with a system
    \item \textbf{Honest}: true, as comprehensive as necessary, and without intent to deceive by system builders or owners
\end{itemize}

In this paper, we provide a \textbf{condensed overview of the diverse conceptualizations of \emph{transparency} in the AI literature}, 
identify commonalities and differences among them,
and discuss how each ties in to our transparency ideal.
We identify three overarching \emph{factors} with which
transparency is invoked concerning the machine learning pipeline---data (\cref{sec:input}), systems (\cref{sec:system}), and outputs (\cref{sec:output}).
We divide our literature review into sections based on these factors and identify specific \emph{clusters} of thematically related research. 
For each cluster, we summarize the high-level issues it approaches, 
briefly detail a representative study,
and provide remarks on its promise and obstacles to advancing the high-level goal of transparency in AI.
We conclude by discussing commonalities and conflicts between the factors and clusters and meditate on the role transparency research will play in a world increasingly dominated by AI systems and services (\cref{sec:commonthreads}).



\section{Data-Related Transparency Factors}\label{sec:input}
One key thrust for transparency work centers on the \emph{inputs} required to produce an AI system.
These studies focus on 
the intent behind \cite{gebru2018datasheets},
composition of \cite{weissgerber2016static}, 
or use limitations for \cite{bertino2019data} datasets as well as 
address the conflicts that can arise 
between transparency and user concerns about data privacy and security
\cite{jordan2020transparency,beam2020challenges}. 
Research toward these factors often explores ways to strike
a balance between increasing overall transparency 
to reap the attendant benefits regarding fairness, accountability, and trust,
while mitigating the potential losses 
vis-a-vis privacy and security.

For our analysis of data-related transparency, we distinguish between works focused on 
information about \emph{training data used to produce models} (\cref{subsec:data}), and
about the \emph{active use of user data by a system} (\cref{subsec:privacy}) in the course of its operation.

\subsection{Transparency on Model Training Data}\label{subsec:data}
The behaviors of machine learning systems fundamentally follow from 
the nature of their training data. 
Information about model training data is thus integral to addressing fairness concerns and ensuring quality \cite{yanisky2019equality}.
Policymakers have begun to mandate disclosures around training data \cite{gdpr} 
and downstream developers and vendors desire understanding of training
dataset limitations \cite{Felzmann2019}
and model-data usage \cite{bhatt2020machine}. 
To reach these desired goals,
\citet{bertino2019redefining} introduce
the terms \emph{record/use transparency} 
as well as \emph{disclosure/data-provisioning transparency}.
While both involve assessing the limitations of training 
datasets, the former is oriented toward holistic quality in AI systems, while the
latter focuses more narrowly on issues of data misuse.

Record transparency is achieved
by describing datasets
with enough contextual information 
for developers to understand how to use them. 
Use transparency---defined as communicating the 
specific purposes for which a dataset is appropriate---often 
complements with record transparency.
For example, \textit{datasheets for datasets} provide 
both record and use transparency to developers when the data provider 
details the production 
and intended use for a resource \cite{gebru2018datasheets}. 
Other studies have improved upon these fact sheets with interviews \cite{hind2020experiences} 
and outlining dataset production best practices that enable more effective and comprehensive documentation \cite{hutchinson2021towards}.

In terms of disclosure/data-provisioning transparency, previous works have found disambiguation of terminology, visualization, and logging systems \cite{bertino2019redefining} particularly useful. These studies claim that these efforts can help unite researchers using the data under a unifying terminology \cite{calgua2022covid} and protected consumer groups (e.g., children) \cite{milkaite2020child} better understand the data process, which in turn provides for better data transparency.

\paragraph{Our view.} Dataset datasheets and other associated record transparency techniques are useful 
for our core transparency goals, insofar as they enable downstream developers and system providers to more 
\emph{honestly} describe the conditions under which their system was produced. Furthermore, along with 
data provisioning transparency techniques the proper \textit{social situatedness} of systems can be ensured, 
as behaviors including differential performance across protected classes or ingestion of data from protected 
consumer groups can be accounted for prior to deployment.

However, 
strong rules and norms that incentivize system developers and providers to actually implement honest and 
socially situated transparency are needed to ensure that this data information leads to ideal AI transparency for users.

\subsection{Transparency on the Handling of User Data}\label{subsec:privacy}
Most if not all useful AI systems must ingest some user data to function.
Demand for transparency around the use of this data is natural considering the privacy and security implications.
While many consider entities who collect, process, store, or train models on user data responsible for respecting user privacy and ensuring secure data handling \cite{camp2015respecting}, 
the specifics of sufficient responsibility or due diligence 
tends to be underspecified
\cite{wieringa2021data}.

Data policy in the US (and other jurisdictions) is largely unregulated, providing industry with free reign \cite{amos2021privacy}. 
Consequently, in unregulated territories, the common practice is to use standard privacy notices written in legal jargon, offering users the option to agree or decline. 
On one side, users may feel forced to accept policies without understanding or contesting them due to a lack of alternatives. \cite{Jensen2004PrivacyPA}.
Contrarily, unaware that this disclosure only appears transparent, users may falsely believe they have control.
\cite{acquisti2013gone}.
By contrast, the EU has taken
a more active approach, passing 
legislation concerning data policy
and consumer privacy.
Last decade, the \cite{gdpr} introduces the \textit{GDPR},
and among other concerns,
demands ``more comprehensible information to end users''
in applicable regions. These demands form the basis of data governance, but require more clarity and precision \cite{grunewald2021tilt}. 

Despite user privacy calls remaining gray, 
calls for data protection of consensually collected data are more concrete.
Classic security research findings are directly applicable in this domain
\cite{kantarcioglu2019securing,naucke2019homomorphically}.
However, the assurance of data protection to stakeholders by providers
is a separate problem. 
A simple solution is to provide a standardized
checklist answerable to the common user 
when transparency of data is clearly communicated \cite{laoutaris2018data}.
Examples of questions include 
``is it leaking to other unintended recipients?'' 
and ``what are the consequences of such leakage?'' 
Norms around answering such questions motivate developers to mitigate identifiable risks.

In addition to legal compliance, companies often address consumer privacy and security concerns to win consumer favor \cite{morey2015customer}. To achieve this, clarity and precision in disclosure in a manner that does not harm privacy and security is necessary \cite{firmani2019ethical}.

\paragraph{Our view.} Ensuring the privacy and security of user data is a core \emph{user-centered} requirement. 
As privacy and security are generally desired, providers have a natural incentive to assure users of their protection.
This can lead to natural tensions with \emph{user-appropriateness} and \emph{honesty} concerns. 
Complex pipelines can have many points of failure, and selling user data is often a profit center for system providers.
Given this conflict between user desires and business realities, why provide true privacy when you can lie? 
Resolving this tension may require strong societal norms and regulation.

When new rights around user data access, understanding,
and protection such as those in GDPR are granted, 
\emph{transparency tools} that actually enable users to exercise these rights must follow.
Producing them is an open technical question in itself \cite{hedbom2008survey}.
Hedbom identifies these as requiring both system-level insight (\cref{subsec:explainability}, \cref{subsec:rationales})
and an ability to understand decision-level modifications required to change output behavior (\cref{subsec:reproducibility}).

\section{System-Centered Transparency Factors}\label{sec:system}
We consider here any work toward elucidating the functionality and quality of AI systems---including methods directed at both practitioners and users.
Practitioners often need to debug models 
or reproduce results more easily \cite{beam2020challenges}. 
On the other hand, users tend to simply desire a basic
overview of a system's function for confidence in its functionality  \cite{mei2022mitigating} (\cref{subsec:disclosive}).
Many ML systems are \textit{black boxes}, providing no insight into the connection between input and output. 
This fundamental lack of functional transparency hinders the \textit{explainability} (\cref{subsec:explainability}) of the system's downstream outcomes \cite{adadi2018peeking,doran2017does}. 
Neural networks, quintessential black boxes \cite{castelvecchi2016can}, are so dominant in AI research that papers claiming to ``open the black box'' have been steadily published for at least 20 years \cite{dayhoff2001artificial}. 
More recently, \textit{automated rationale generation} (\cref{subsec:rationales}) from model-internal states has also grown in popularity \cite{ehsan2019rationalization}.

\subsection{System Function Disclosure}\label{subsec:disclosive}
\emph{System function disclosure} includes 
communications by system producers, owners, or vendors concerning 
the capabilities and limitations of their systems. 
A challenge in making prescriptions around this sort of transparency is that
system function disclosures target a diverse set of audiences, including
external developers building around/needing to understand a system \cite{Felzmann2019,vaughan2020human},
lay users of a system \cite{saxon2021modeling},
or regulatory bodies \cite{ec2019ethics}. 
The Association for Computing Machinery (ACM) even 
considers this sort of disclosure required in its Code of Ethics \cite{gotterbarn2018acm}. 

Frameworks for concise communication of model strengths and limitations are instrumental to effective system function disclosure. For example, \emph{model cards} 
provide a simple set of data points for developers to communicate the limits and intended use-cases of their models \cite{mitchell2019model}. 
However, prescribing that disclosure takes place doesn't ensure that the disclosure contains
useful information,
or the information provided will be relevant and understandable to those consuming it. 
To address this issue, work on 
qualitatively evaluate the disclosure sufficiency with rubrics \cite{barclay2019quantifying,barclay2021framework}, or automatically assessing the layperson-comprehensibility of a system function description \cite{saxon2021modeling} have been proposed.

A further challenge to system function disclosure is that, in many cases, even experts don't know precisely how black
box systems mechanistically produce output from input \cite{rudin2019stop}---thus explainability
and interpretability techniques can be prerequisites for the level of expert understanding needed to produce \emph{honest} 
disclosure (\cref{subsec:explainability}).

A common limitation to many ML techniques applied in sensitive settings (e.g., medicine, criminal justice, employment) is the invisibility of external social context to the model.
As biases and oversights in training data propagate to learned systems, system developers require clarity
from data providers (\cref{subsec:data}) to ensure that they can in turn communicate the problems of their systems \cite{gebru2018datasheets}.
Furthermore, in these complex settings systems often lack the sort of commonsense knowledge that is required
to effectively operate in a human-centered environment \cite{riedl2019human}. 
Apart from solving the problem of commonsense reasoning in AI, actively soliciting direct user feedback
to contextualize failure cases \cite{suzor2019we} is one practical way to bootstrap documentation of model weaknesses. 

However, even if perfect information about a model's strengths and weaknesses (which is difficult to gain)
and strong explanations of its internal functional details are available, calibrating explanations to be
comprehensible to diverse stakeholders is still a confounding problem. 
In short, different groups require different explanations, and have different levels of expertise. 
Insufficient disclosure may be unsatisfying, but too much disclosure may result in 
information overload, and lower user trust \cite{knowles2017intelligibility}. 
To combat this issue, it is crucial to understand the desired ends of each stakeholder and 
information in a manner that balances this duality \cite{vaughan2020human}.   

\paragraph{Our view.}
At its best, system function disclosure advances the goals of both \emph{user-centric} and \emph{honest} communication. 
Users who understand how a system works are empowered to make their own decisions regarding it. However, ensuring that
these communications are \emph{user-appropriate} is particularly challenging, as considerable expertise is involved
in producing the systems, and they sometimes aren't easily reduced to layperson-appropriate explanations \cite{xu2021self}.

Furthermore, enhanced system-level transparency may introduce security risks, as information about the function of a system
can be utilized by prospective attackers \cite{jordan2020transparency}. Balancing the needs of disclosure and security must
be performed carefully---we hope future work will guide norms and regulations toward such a balance.

\subsection{Explainable AI and Causality}\label{subsec:explainability}
This section discusses transparency through information provided by systems, rather than human disclosure (\cref{subsec:disclosive}). We will focus on the connection between \textit{explainable AI} techniques and transparency, rather than a complete overview.

Many simple ML models, such as decision trees or support vector machines are fundamentally, casually explainable \cite{rudin2019stop}. 
However, these simple models lack the flexibility of 
opaque neural networks \cite{castelvecchi2016can}. 
Some attempts to render neural nets more interpretable focus on converting their massive inscrutable internal weight matrices into something simpler, such as training under constraints like forced sparsity \cite{du2019techniques}, or distillation to an explainable student model (such as a simpler linear classifier) whose outputs can then be analyzed \cite{tan2018distill}. 

Other methods instead directly peek inside the black box. 
Some neural net architectures, such as attention mechanisms, 
are often presented as being fundamentally interpretable due to their easy generation of salience maps which provide insight into the output correlation of input features \cite{ribeiro2016why,hendricks2016generating}. However, it is debated whether these maps provide any explanatory or actionable insight into how these architectures actually operate \cite{jain2019attention,wiegreffe2019attention}. Due to their poor intelligibility to end-users these ``explanations'' can even lower user trust for a system \cite{schmidt2020transparency}. 

Input influence methods are often positioned as explainability techniques. 
Influence functions to interpret input variations \cite{koh2017understanding} and quantitative measures to capture an input's degree of influence \cite{datta2016algorithmic} 
have diverged from the causality interpretation \cite{hamon2021impossible} of good explanations. 
Removal of all confounding variables from natural datasets is realistic-to-impossible. 
Thus, models trained on natural data---including naturally interpretable regression models---will not reflect a causal relationship.
Doing so requires identifying a backdoor adjustment variable set which, when conditioned on, guarantees causality by eliminating all confounders \cite{pearl2009causality}. 
Only when these variables are conditioned upon can we assume that a statistical correlation does imply causation. 
Rarely is this condition satisfied.

Explainability through \textit{counterfactual reasoning} \cite{wachter2017counterfactual}, 
leverages
counterfactuals to explain what inputs achieve desired outputs. Without altering black-box models, this is an effective strategy that uses propositional logic to provide interpretable reasoning to users, where they can decide whether a decision is trustworthy. For example, if the reasoning were based on a protected variable, it would be obvious the machine is discriminating. However, if the reason were poor credit history for a loan application, the decision would be reasonably sound and trusted by the user. 

\paragraph{Our view.}  Strevens argues that a good explanation answers a ``why'' question \cite{strevens2011depth}.
We are inclined to agree, and believe that explanations establishing a true causal interpretation
best advance the ideals of \emph{honesty} and \emph{user-centricity}, as a causal explanation empowers 
users with understanding of how their chosen inputs affect outputs.
However, with increasing complexity of systems, producing \emph{user-comprehensible} explanations grows ever more challenging.
We hope for further work to improve this state of affairs.

We view ``explanations'' grounded in non-causal relationships such as feature maps or influence functions to be of 
dubious honesty to end-users. While they are useful for expert analysis and debugging, they
could be persuasively employed to trick critics into trusting a system in which trust is not deserved. 
As \emph{automation bias} is generally an issue with the organizational deployment of automated systems, 
care must be taken to ensure that \emph{socially situated} and properly contextualized explanations be given to users to ensure trustworthiness \cite{ehsan2021expanding}.
Thus, it is imperative that policymakers receive clear messaging from the research community on the strengths and limitations of
explainable AI systems.

\subsection{Generated Rationales}\label{subsec:rationales}
\cite{ehsan2019rationalization} introduce \textit{automated rationale generation}, an alternative form of explainability that seeks to map a model's internal state into human-interpretable rationales in natural language. While these generated rationales are not guaranteed through causality to be correct, they provide insight into language models' reasoning abilities \cite{rajani2019explain}. 
These rationales can help users fact-check outputs to mitigate the potential for misinformation \cite{mei2022foveate}. 

Multiple failure modes exist for these rationales, including \emph{hallucination} by the natural language generating component,
which can lead models to provide rationales that differ from the system's decision. 
\citet{jacovi2020towards} provides a survey on faithfulness, defined to capture the community consensus on measuring the hallucination of explanations. Namely, explanations are unfaithful if either of these conditions is satisfied: an explanation does not match the decision, or two explanations differ for similar inputs and outputs. Unfaithful models can contribute to unintentional deception and detract from user confidence. As a result, several studies focus on improving model faithfulness \cite{zhang2021sample,lakkaraju2019faithful}.   

\paragraph{Our view.} 
Although they risk incorrectness, these rationales have the potential to further the
\textit{user-appropriateness} and \textit{user-centricity} ideals. Such research is fundamentally oriented toward
providing user-interpretable rationales.
Should the evaluation metrics align with human comprehensibility, the generations rationales are style-appropriate. 

However, extreme care must be taken in crafting norms around these systems to ensure that \emph{honesty} is centered.
After all, the ability for a system to generate some rationale for its decision is no guarantee of its accuracy.
Future research to resolve this issue might take the form of some kind of higher-level fact checking to ensure that 
these rationales are true, but evaluating NLG explanations is a challenge \cite{clinciu2019survey}. 
Lay users must be educated on the degree of trust that these systems deserve to mitigate these risks.

\section{Output-Oriented Transparency Factors}\label{sec:output}
\textit{Output-oriented} transparency is directed toward ensuring sufficient system performance for stakeholders.
This thread of research distinguishes how similar concepts in the system demonstration space are differentiated (\cref{subsec:reproducibility}),
such as \textit{repeatability}, \textit{replicability}, and \textit{reproducibility} \cite{acm2016reproduce} as well as exact, empirical, and conceptual reproducibility \cite{aguinis2019transparency}. 
Studies in this direction 
face problems around the \emph{degree} of transparency disclosure, 
weighing competing considerations
of providing sufficient information about a system to a stakeholder 
without overloading information \cite{pieters2011explanation,saxon2021modeling}. These discussions about system demonstration are often motivated by a desire for fairness, accountability, or trust \cite{vaughan2020human,Felzmann2019}, which are often positively associated with increased transparency (\cref{subsec:fairness}). Explainable models can promote fairness and accountability, but only if properly aligned. Misaligned explanations can harm privacy and security \cite{strobel2019aspects,shokri2021privacy} (\cref{subsec:privacy}).

\subsection{System Demonstrability}\label{subsec:reproducibility}
System demonstration is necessary to support claims of function, performance (\cref{subsec:disclosive}), consistency, privacy, and security (\cref{subsec:privacy}).
Works within this cluster typically concern \textit{repeatability}, \textit{replicability}, or \textit{reproducibility} as defined by the ACM \cite{mora2021traceability}. 
Of these notions, repeatability is the easiest to articulate---repeatable systems and methods produce the same outputs over the same inputs and experimental setup;
this basic result consistency is generally assumed and not further discussed \cite{duggin1990assumptions}. 
The other two notions, replicability and reproducibility receive more attention in the AI transparency literature \cite{beg2021using,lucic2021reproducibility}. Both these forms require a different team to achieve the same system performance with the same setup (replicability) or a different setup (reproducibility).

Technologies including cloud storage and compute services, 
environment management systems, 
and interactive multimedia/code documents \cite{vogtlin2020cobra}, 
can all enable more replicable research. 
Relying solely on these technologies to ensure replicability has limitations.
Persisting cloud instances or sharing machine images can be costly, difficult, or against the policies of research entities \cite{clyburne2019computational}. 
Packages may have unstable dependencies and sit outside of public package repositories, hindering the utility of automated environment management systems. 
Identical experimental setups may be difficult due to software versioning or physical hardware. 
Jupyter notebooks and other combined media/code documents are vulnerable to these issues; additionally, they can introduce problematic user interface factors that further hinder replicability, such as unclear order of operations. 
Moreover, even if all hurdles to technical replicability are overcome, the overall reproducibility of an experiment may be preempted by non-technical factors.

ML study reproducibility can require varying levels of strictness and precision, such as exact, empirical, or conceptual reproduction \cite{mora2021traceability,aguinis2019transparency}. 
As the procedure abstracts conceptually, it becomes harder to draw the same conclusions empirically, thus increasing robustness. 
For example, procedures using different hyper-parameters and achieving similar results can show that a system is robust \cite{brendel2019accurate}. 
As researchers strive for conceptual reproducibility of their work, the credibility and robustness of the concepts, models, and ideas will increase as the ability to demonstrate a model's performance directly relates to the certainty of its performance.


\paragraph{Our view.} The distinctions of repeatability, replicability, and reproducibility, as well as the types of reproducibility encourage \textit{honesty} from developers. Such terminology fosters an atmosphere of precise communication that mitigates confusion and deception from the  potential for terminology overloading, thereby also encouraging robust AI systems. Additionally, this research cluster heavily intertwines with \textit{user-appropriateness}. In the optimal setting, disclosure regarding system demonstrability should be granular enough to allow duplication of results, while mitigating unnecessary information. However, in practice, the context in which this disclosure is conveyed is important as these details could be instead used as an information overload tactic to deceive downstream users into trusting such a system by touting robustness, when in reality such a user would not find this information meaningful nor likely have the resources to duplicate such results. To this end, we believe these efforts should empowers stakeholders with more helpful information to decide whether adapting such AI systems make sense for their respective use case, but are not necessarily relevant for the downstream user.

\subsection{Fairness and Accountability}\label{subsec:fairness}
Transparency is crucial for ensuring fairness. Lack of transparency raises fairness concerns and undermines trust in AI systems. Inaccurate results can come from two forms: system fragility and systematic bias. Fragile systems are poor quality applications that need to be validated to ensure reproducibility, for general usability (\cref{subsec:reproducibility}). Once a system is adequately robust to technical bugs, systems may still be systematically biased, where decisions are unfair toward some groups, raising eyebrows for regulators. 
Similarly, negative decisions for users lead to dissatisfaction and a desire for interpretable explanations. 
With uninterpretable decisions, we raise the same concerns \cite{toreini2020relationship}. An application outside of AI occurs in content moderation, which often lacks transparency in moderating decisions made (i.e., suspension) \cite{suzor2019we}. The lack of explainability often reduces consumer and public trust \cite{toreini2020relationship} for fairness \cite{veale2018fairness} and accuracy \cite{mcsherry2005explanation}.
Explainable models increase accountability for fairness concerns.


While unfair decisions to a consumer's detriment are a focus, \cite{wang2020factors} introduces \textit{favorability bias}, where users perceive a system decision as fair when that decision is beneficial. Beneficial decisions are skewed in favor of a trusted, fair decision, while unfavorable decisions are the contrary \cite{liaotrust2022,springer2018hiding}. Thus, a call for both interpretability and human disclosure is needed to alleviate these concerns.

\paragraph{Our view.} Work in this cluster encourages \textit{honesty} from systems and developers. A proactive stance on fairness encourages 
increasingly transparent and explainable models, to allow for accountability. Naturally, fairness involves \textit{social context} as accountability is desired to mitigate systemic bias.

\begin{table*}[t!]
\centering
\resizebox{\linewidth}{!}{
\begin{tabular}{ l p{1.2\linewidth}} 
 \toprule
 \textbf{Stakeholder} & \textbf{Selected desired ends.} \\ 
  
  \midrule
  Deployer & lead a user into some action or behavior, increase usage of their system, maintain a functional system \\
  
  \midrule
  Developer & understand a system to debug and improve it, predict real-world system behavior, improve system performance and robustness  \\
  
  \midrule
  Data Owner & provide data collection and usage information, protect proprietary data and trade secrets, address data misuse concerns \\ 

  \midrule
  Regulator & evaluate fairness of predictions, demonstrate regulatory compliance, managing societal risk, mitigating negative consequences \\ 

  \midrule
  User & understand system logic, evaluate trustworthiness, recognize AI model's socioeconomic blindspots, data protection and privacy \\

  \midrule
  Society & understand the strengths and limitations of a system, overcome fear of the unknown, encouraging ethical use of AI, mitigating system bias  \\
  
  \bottomrule
\end{tabular}
}
\vspace{-1ex}
\caption{A selection of stakeholders and their various desired ends relating to AI transparency.}
\label{table:stakeholders}
\end{table*}

\section{Discussion}\label{sec:commonthreads}

In spite of their sometimes contradictory goals, each of the aforementioned clusters has a role to
play in realizing our ideal vision for AI transparency.
We conclude by discussing common across the overarching factors and clusters of transparency research,
motivated around \textit{study attributes} where these commonalities and conflicts play out. In particular, we discuss how 
\textbf{desired ends}, \textbf{associated stakeholders}, and \textbf{utilized means} relate and differ across them, and how these 
attributes can either advance or hinder our ideal of \emph{user-appropriate}, \emph{user-centered}, and \emph{honest} AI transparency.


\subsection{Desired Ends}\label{subsec:ends}




Different stakeholders in transparency work have different desired ends \cite{weller2017challenges,Felzmann2019,vaughan2020human,bhatt2020machine}. 
However, many of these desired ends fundamentally conflict---
\textbf{no means exist that can simultaneously satisfy all stakeholders desires}. 
This is complicated by the fact that there is a lopsided power dynamic between 
the \textit{empowered stakeholders} (i.e., owners, developers, and deployers) 
who choose means of transparency and the lay users who cannot (\autoref{table:stakeholders}).

For example, developers may seek explainability to debug a system \cite{doran2017does} or foster end user trust \cite{balog2019transparent} (\cref{subsec:explainability}). Legislators may invoke transparency requirements to enable visibility by regulators \cite{castillo2019fairness} or to drive more equitable outcomes across demographic groups \cite{sweeney2019transparent} (\cref{subsec:fairness}). However, in other cases studies' desired ends lie in the margins.

Orthogonal to transparency about or from AI systems is transparency regarding the broader context in which
they're deployed. Often, this is a question of clearly stating the goals with which a system, such as a social media content moderation
pipeline, is deployed \cite{andrada2022varieties}. While providing transparency on how an AI system implementing some goal functions is 
instrumental to giving comprehensive transparency around a sociotechnical system, information on the overall goals of the system (e.g., 
corporation, website, or platform) is necessary to give users an \emph{honest} understanding.


\subsection{Conflicting Means}\label{subsec:means}

\begin{table*}[t!]
\centering
\resizebox{\linewidth}{!}{
\begin{tabular}{ l p{1.2\linewidth}} 
 \toprule
 \textbf{Means} & \textbf{Criteria for such means.} \\ 
  
  \midrule
  Human Disclosure & information provided by humans to improve clarity in understanding an AI system (i.e., disclosure of dataset demographics as social situatedness) \\
  
  \midrule
  System Disclosure & information outputted from systems to improve clarity in understanding of the system (i.e., disclosure of generated rationales for human intelligibility)  \\
  
  \midrule
  Deception & disclosure of content that intentionally or unintentionally misleads (i.e., dishonest disclosure to tout system performance) \\ 

  \midrule
  Info. Overload & disclosure of a surplus of information that overwhelms (i.e., providing hyper-parameters to users as substitute for user-appropriate information) \\ 
  
  \bottomrule
\end{tabular}
}
\vspace{-1ex}
\caption{Means for transparency: human/system disclosure positively contribute, while deception/information overload negatively contribute.}
\label{table:means}
\end{table*}

Among the many means that may be employed to achieve the various ends of transparency (\cref{subsec:ends}) 
we identify \textit{human disclosure}, \textit{system disclosure}, \textit{deception}, and \textit{information overload} as four
techniques that appear or are discussed in the literature (\autoref{table:means}). In the ideal case, 
human- and system-disclosure trivially achieve many transparency ends desired by all stakeholders. However, these techniques can conflict with the intellectual property protection 
needs of system developers, deployers, and owners. Furthermore, these empowered stakeholders can mislead the less 
empowered ones, either deliberately through \textit{deception} or inadvertently via \textit{information overload} \cite{poursabzi2018manipulating}. This conflict is particularly problematic for producing effective regulation. 
Deceptive appeals to transparency can constitute a form of ``ethics washing'' \cite{yeung2019ai,bietti2020ethics} 
wherein empowered stakeholders use the veneer of ethics to shield against regulatory scrutiny \cite{wagner2018ethics} 
or build potentially unearned user trust \cite{benkler2019don}, but even well-meaning empowered actors can confuse 
users under an information overload (\cref{subsec:explainability}, \cref{subsec:reproducibility}).

An important note that while the explanation as system disclosure may ultimately be the same in both cases, and thus be able to achieve the same desired end such as to ``lead a user into some action or behavior'' for a deployer, this example causes a conflict with the user's desired to ``determine the trustworthiness of a system.'' 

For example, much work in the data- (\cref{subsec:data}, \cref{subsec:privacy}) and system-oriented (\cref{subsec:disclosive}) communicative domain 
centers producing norms and standards around disclosures, rather than leaving developers and vendors to produce ad-hoc standards. However, the needs and risks of transparency aren't standard in AI as they are in fields that inspired such disclosures 
(e.g., electronic components \cite{gebru2018datasheets}).

\subsection{Selection of Study Attributes}
\paragraph{Researcher Interest.} There is a fundamental tension between financial results and responsible AI requirements in corporate settings. Businesses may favor reducing legal liability and increasing their competitive edge \cite{chen2021fairness}. Additionally, research can be heavily impacted short-term and long-term business needs. Yet, authors positioned within entities that maintain large AI systems are often uniquely positioned to assess their real-world impacts \cite{pfau2020case}. It would be a mistake to discount work produced by interested industrial parties wholesale.

Furthermore, the sometimes wide-reaching societal impacts of the AI systems under study by researchers may lead them to act as interested stakeholders in society at large. For example, OpenAI researchers cites the dangers of large language model abuse by 
malicious actors with long-term agendas as a motivation for their tiered-release strategy of GPT-2 \cite{solaiman2019release}.
OpenAI's subsequent GPT-3 model was never open-sourced, with the firm instead opting to control and sell API access to the model 
\cite{vincent2021microsoft}.

Dealing with researcher interest in evaluating transparency work is thus a balancing act. Motivations should be carefully considered, particularly when self-interest might conflict with desirable research goals, which researchers are disincentivized to disclose. These considerations must be balanced with good-faith reading, and the understanding that everyone has a stake in the societal impact of AI system. 


\paragraph{Persuasion vs. Trust.}

Measures of effectiveness in transparency varies considerably on the target measure. Is the goal really to empower users, or simply to assuage their concerns? Navigating which is in play in a given study can be a challenge.
The definition of an explanation is debatable \cite{jain2019attention}, 
explanations often do not provide a causal justification \cite{krishnan2020against}. 

A non-causal explanation is not actionable; for this reason we believe causality in explanations to
be core to achieving both \emph{user-centered} and \emph{honest} transparency (\cref{subsec:explainability}). 
How are users to understand this? There are no easy answers to this question.
The incentive is to persuade users to believe in systems, regardless of if a disclosure is honest or not (\cref{subsec:data}, \cref{subsec:disclosive}). 
This is why \emph{user-appropriate} education on the function of AI systems is so crucial---to empower them 
to evaluate for themselves whether the \textbf{claims} and \textbf{actions} of providers align.

\paragraph{Claims versus Actions.} 
A complicating factor in evaluating research on transparency in AI is that often the stated goals or claims of the researchers differ from their actions. 
Consider the persuasion versus trust dichotomy, which represents how work claiming to address ``trust'' toward systems by convincing prospective users to use it rather than by demonstrating the trustworthiness of the system.
Works claiming to provide explanations of a system decision instead may only provide tangential information about the decision \cite{yeung2018how}. For example, is an attention map that lead to a text classification outcome---often touted as an explanation (\cref{subsec:explainability})---really explaining the decision? Across multiple disciplines, the answer is probably no. 

For legal scholars, a good explanation must be appropriate for the recipient of the explanation \cite{raz2011normativity}, written in language understandable to them (\cref{subsec:rationales}). This is clearly reflected in the language of the EU guidelines for ethical AI, in which explanations are to be ``appropriate to the end-user at hand'' \cite{ec2019ethics}. Both legal scholars and counterfactualists require a but-for distinction to be made for an explanation to fly \cite{yeung2018how}.

Current proposals around AI transparency, particularly the European Commission Guidelines for Trustworthy AI \cite{ec2019ethics}, are steeped in AI research. Furthermore, scientific researchers frequently claim or insinuate that they are producing research work that aligns with policymaker claims or interests.
As a community, we should conduct ourselves with this downstream impact on major societal actors in mind. In particular, this means \textbf{precision in terminology and alignment of claims with reality in systems must be prioritized in work and enforced in peer review}.



\section{Conclusion}

We have overviewed a broad sample of AI transparency-related works, 
situated them based on overarching factors into narrower cluster,
identified their common threads, 
and tied them into our vision for user-appropriate, user-centered, honest communication.
We welcome further negotiation regarding what specific fundamental attributes characterize these transparency studies. 
We hope that this direction can lead to productive scientific and regulatory advancements \cite{wischmeyer2020artificial}, 
and ultimately enable a more precise public discourse.
This ontology represents a first step toward resolving the problem of terminological 
imprecision of regarding system transparency. 

As conferences continue to ask more of authors in reproducibility and ethical considerations \cite{gotterbarn2018acm}, 
we would like to see movement on asking for specificity in the terminology around transparency to reduce gray areas for improperly aligned works.
The research community must produce a body of work from which lawmakers, practitioners, and the general public can clearly understand 
\textit{what transparency means for AI systems}. 
To the extent legislation follows academics and corporate actors, it is crucial that the community speak honestly, clearly, and unambiguously. 

\textbf{Specificity in what is meant by \textit{transparency} as it is used must become a norm in AI research communities.} To this end, we would like to see further advancement in characterizing the attributes of transparency studies, discussing the means-ends tension, the stakeholders behind and engaged in the study, and the use of these frames in a systematic manner to contextualize claims, techniques, and results across the human-concerned and system-derived transparency spaces. Ultimately, taking these proactive actions will give users and stakeholders the resources to confidently decide whether to trust a particular AI system.



\appendix





\scriptsize
\bibliographystyle{named}
\bibliography{ijcai23}

\end{document}